\documentclass{iopart}
\usepackage{iopams,psfig,graphicx}
\begin{document}
%%%%%%%%%%%%%%%%%%%%%%%%
\title{Nonlocality of two- and three-mode continuous variable systems}
%%%%%%%%%%%%%%%%%%%%%%%%
\author{Alessandro Ferraro and Matteo G. A. Paris}
\address{Dipartimento di Fisica and Unit\`a INFM,
Universit\`a di Milano, Italy}
%%%%%%%%%%%%%%%%%%%%%%%%%%%%%%%%%%%%%%%%%%%%%%%%%%%%%%%%%%%%%%%%%%
\begin{abstract}
We address nonlocality of a class of fully inseparable three-mode
Gaussian states generated either by bilinear three-mode Hamiltonians or
by a sequence of bilinear two-mode Hamiltonians. Two different tests
revealing strong nonlocality are considered, in which the dichotomic
Bell operator is represented by displaced parity and by pseudospin
operator respectively. Three-mode states are also considered as a
conditional source of two-mode non Gaussian states, whose nonlocality
properties are analyzed. We found that the non Gaussian character of the
conditional states allows violation of Bell's inequalities (by parity
and pseudospin tests) stronger than with a conventional twin-beam state.
However, the non Gaussian character is not sufficient to reveal
nonlocality through a dichotomized quadrature measurement strategy.
\end{abstract}
%%%%%%%%%%%%%%%%%%%%%%%%
\date{\today}
%%%%%%%%%%%%%%%%%%%%%%%%
\section{Introduction}\label{s:intro}
Einstein, Podolsky and Rosen (EPR) formulated their famous argument
about the completness of quantum mechanics in the framework of
continuous variable systems \cite{EPR}. However, after that Bohm gave a
dichotomized version of it \cite{Bohm}, the debate concerning
nonlocality moved to systems described by discrete variables, leading
Bell to formulate his celebrated inequalities in a dichotomized fashion
\cite{Bell}.  Recently, the increasing importance of continuous variable
systems leads many authors to explore the nonlocality issue in its
original setting, where dichotomic observables to test Bell's
inequalities are not uniquely determined. The attempts to translate
Bell's inequalities to continuous variable systems clarified the fact
that crucial in a nonlocality test is the existence of a set of
dichotomized bounded observables used to perform the test itself, from
which the so called ``Bell operator'' is derived. The more debated
question has dealt with the nonlocality of the normalized version of the
original EPR state, {\em i.e.} the twin-beam (TWB) state of radiation
produced by spontaneuos downconversion in a parametric amplifier
\cite{WallsMilburn}. Nonlocality of the TWB state was not clear for a
long time. Using the Wigner function approach, Bell argued that the
original EPR state, and as a consequence the TWB too, does not exhibit
nonlocality because its Wigner function is positive, and therefore
represents a local hidden variable description \cite{BellSpkUnspk}. More
recently, Banaszek and Wodkiewicz \cite{BW} showed instead how to reveal
nonlocality of the EPR state through the measurement of displaced parity
operator. Furthermore a subsequent work of Chen et al. \cite{PS} showed
that TWB's violation of Bell's inequalities may achieve the maximum value
admitted by quantum mechanics upon a suitable choice of the measured
observables. Indeed, the amount of violation crucially depends on the
kind of Bell operator adopted in the analysis, ranging from no violation to
maximal violation for the same (entangled) quantum state.
\par
Systems which involves only two parties are the simplest setting where
to study violation of local realism in quantum mechanics. A more
complex scenario arises if multipartite systems are considered.
Studying the peculiar quantum features of these systems is worthwhile
in view of their relevance in the development of quantum communication
technology, {\em e.g.}~to manipulate and distribute
information in a quantum communication network
\cite{VLBTelenet,YOF04}. Although the study of multipartite
nonlocality has originated without the use of inequalities \cite{GHZ},
an approach to derive Bell inequalities has been developed
\cite{Mermin_Klyshko} also for these systems and applied to
characterize their entanglement properties \cite{GBP98}. Being originally
developed in the framework of discrete variables, these multiparty
Bell inequalities have found application also in the characterization
of continuous variable systems \cite{VLB,CZ02}. 
\par 
The aim of this paper is to apply the various approaches hitherto
developed to test nonlocality of two- and three-mode continuous variable 
systems. We will consider tripartite Gaussian states as well as
non-Gaussian bipartite states. In the first case strong violation of
Bell inequalities is found, allowing the Bell factor to reach values
of ${\cal B} \simeq 3$, while in the second case enhancement of
nonlocality is obtained in comparison with the TWB case.  
\par
The paper is organized as follows. In Sec. \ref{s:rev} we review the
different approaches to test nonlocality in the framework of
continuous variables and introduce notation that will be used
throughout the paper. The three-mode states we are interested in are
introduced in Sec. \ref{s:3m}, and their violation of local realism 
is analyzed in \ref{ss:DP3} and \ref{ss:PS3}. In Sec.
\ref{s:2m}, the tripartite states are considered as sources for
conditional generation of non-Gaussian bipartite states, whose 
nonlocal proprties are then studied in \ref{ss:DP2}, \ref{ss:PS2} 
and \ref{ss:H2}. Finally, the main results obtained are summarized 
in Sec.~\ref{s:conclusions}, which closes the paper with some concluding
remarks.
%%%%%%%%%%%%%%%%%%%%%%%%
\section{Nonlocality tests for continuous variables}\label{s:rev} 
\par
In this section we will briefly recall the inequalities imposed 
by local realism in the cases of our interest.
Let us start by focusing our attention on a bipartite system. Let 
$m(\alpha_1)=\pm 1$ and $m(\alpha_1')=\pm 1$ denote two possible 
outcomes of two possible measurements on the first subsystem and 
similarly $m(\alpha_2)=\pm 1$ and $m(\alpha_2')=\pm 1$ for the second 
subsystem. The essential feature of this measurements is that they 
are local, dichotomic and bounded. The Bell's combination
\begin{equation}
F_2 \equiv m(\alpha_1)m(\alpha_2)+m(\alpha_1)m(\alpha_2')+
m(\alpha_1')m(\alpha_2)-m(\alpha_1')m(\alpha_2')
\label{F2}
\end{equation} 
under the assumption of local realism gives rise to the well known
Bell-CHSH inequality \cite{CHSH}:
\begin{equation}
{\cal B}_2 \equiv |E(\alpha_1,\alpha_2)+E(\alpha_1,\alpha_2')+
E(\alpha_1',\alpha_2)-E(\alpha_1',\alpha_2')| \le 2
\label{BI2}\;,
\end{equation} 
where $E(\alpha_1,\alpha_2)$ is the correlation function
between the measurement results, \emph{i.e.}, the expectation 
value of the products of the results of the experiments $m(\alpha_1)$ 
and $m(\alpha_2)$.
\par
In the case of a n-partite system, a nonlocality test is possible using
the Bell-Klyshko inequalities \cite{Mermin_Klyshko,GBP98} which provides a generalization 
of inequality (\ref{BI2}). These inequalities are based on the following
recursively defined linear combination:
\begin{equation}
F_n \equiv
\frac12\left[m(\alpha_n)+m(\alpha_n')\right]F_{n-1}+\frac12
\left[m(\alpha_n)-m(\alpha_n')\right]F'_{n-1}
\label{FN}\;,
\end{equation}
where $m(\alpha_n)=\pm 1$ and $m(\alpha_n')=\pm 1$ refer to measurements 
on the n-party of the system, and $F'_n$ denote the same expression as $F_n$
but with all the $\alpha_j$ and $\alpha_j'$ inverted.  In the case of a
three-partite system, local realism assumption imposes the following
inequality from combination (\ref{FN}):
\begin{equation}
\fl {\cal B}_3 \equiv |E(\alpha_1,\alpha_2,\alpha_3')+E(\alpha_1,\alpha_2',
\alpha_3)+E(\alpha_1',\alpha_2,\alpha_3)-E(\alpha_1',\alpha_2',\alpha_3')| \le 2
\label{BI3}\;,
\end{equation}
where again $E(\alpha_1,\alpha_2,\alpha_3)$ is the correlation function 
between the measurement results. Quantum mechanical 
systems can violate inequalities (\ref{BI2}) and(\ref{BI3}) by a maximal
amount given by, respectively, ${\cal B}_2 \le 2\sqrt{2}$ and 
${\cal B}_3 \le 4$ (see, {\em e.g.}, Ref.~\cite{GBP98}).
\par
We now briefly review three different strategies to reveal quantum
nonlocality in the framework of continuous variables systems. Recall
that in the case of a discrete bipartite system, for example a
spin-$\frac12$ two particle system, the local dichotomic bounded
observable usually taken into account is the spin of the particle in
a fixed direction, say ${\bf d}$ . Hence the correlation between two
measurements performed over the two particles is $E({\bf d_1},{\bf
  d_2})=\langle {\bf d_1} \boldsymbol\sigma\otimes {\bf d_2}
\boldsymbol\sigma\rangle$, where the operator $\boldsymbol
\sigma=(\sigma_x,\sigma_y,\sigma_z)$ is decomposed on the Pauli
matrices base and ${\bf d_1},{\bf d_2}$ are two unit vectors. 
The Bell operator is then given by the expression:
\begin{equation}
B_{2,sp} = {\bf d_1}\boldsymbol\sigma\otimes {\bf d_2}
\boldsymbol\sigma+ {\bf d_1'}\boldsymbol\sigma\otimes {\bf d_2}
\boldsymbol\sigma+ \bf {\bf d_1}\boldsymbol\sigma\otimes 
{\bf d_2'}\boldsymbol\sigma- {\bf d_1'}\boldsymbol\sigma
\otimes {\bf d_2'}\boldsymbol\sigma
\label{BO2Spin}\;.
\end{equation} 
\par
Consider now a n-partite continuous variable system identified by the
creation operator $a^\dag_j$ and the annihilation operator $a_j$
($j=1,\dots,n$) with boson commutation relations associated. Following the
original argument by EPR it is quite natural attempting to reveal the
nonlocality of this system trying to infer quadratures of one
subsystem from those of the others. From now on, we will refer to this
procedure as a ``Homodyne nonlocality test'', as quadrature measurements
of radiation field are performed through homodyne detection. Here we
identify the quadrature $x_j(\theta)$ according to the definition
$x_j^\theta=\frac{1}{\sqrt2}(a_je^{-i\theta}+a^\dag_je^{i\theta})$. As
they are local but neither bounded nor dichotomic, quadrature
observables are not immediately suitable to perform a nonlocality test
based on Bell's inequalities. The procedure to make them bounded and
dichotomic is quite arbitrary and consist in the assignment of two
domains $D_+$ and $D_-$ to each observable \cite{Gilchrist}. When the result
of a quadrature measurement falls in the domain $D_\pm$ the value
$\pm1$ is associated to it. Usually the choice $D_\pm=\mathbb{R}^\pm$
is considered, though a choice suitable to the system under
investigation may be preferable. Considering a bipartite system we can
introduce the following quantities:
\begin{eqnarray}
P_{++}(x_1^\theta,x_2^\varphi)=\int_{D_+}dx_1^\theta
\int_{D_+}dx_2^\varphi P(x_1^\theta,x_2^\varphi) \nonumber \\ 
P_{+-}(x_1^\theta,x_2^\varphi)=\int_{D_+}dx_1^\theta
\int_{D_-}dx_2^\varphi P(x_1^\theta,x_2^\varphi) \nonumber \\  
P_{-+}(x_1^\theta,x_2^\varphi)=\int_{D_-}dx_1^\theta
\int_{D_+}dx_2^\varphi P(x_1^\theta,x_2^\varphi) \nonumber \\ 
P_{--}(x_1^\theta,x_2^\varphi)=\int_{D_-}dx_1^\theta
\int_{D_-}dx_2^\varphi P(x_1^\theta,x_2^\varphi) 
\label{PH}\;,
\end{eqnarray} 
where $P(x_1^\theta,x_2^\varphi)$ is the joint probability distribution of
the quadratures $x_1^\theta$ and $x_2^\varphi$. We can now identify the
homodyne correlation function $E_H(\theta,\varphi)$ as
\begin{equation}
E_H(\theta,\varphi)=P_{++}(x_1^\theta,x_2^\varphi)+
P_{--}(x_1^\theta,x_2^\varphi)-P_{+-}(x_1^\theta,x_2^\varphi)
-P_{-+}(x_1^\theta,x_2^\varphi)
\label{EH}\;,
\end{equation} 
which can be straightforwardly used to construct the Bell combination
${\cal B}_{2,H}$ of Eq.(\ref{BI2}) and to perform the
nonlocality test. The main problem of pursuing such a nonlocality test
is that it is not suitable in case of systems described by a positive
Wigner function, as the TWB state of radiation defined as
$|X\rangle= \sqrt{1-X^2}\sum_{n} X^n |n\,n\rangle$, where $X=\tanh r$
and $r$ is the squeezing parameter. Indeed, a positive
Wigner function can be interpreted as a hidden phase-space probability
distribution, preventing violation of Bell-CHSH inequality unless the
measured observables have an unbounded Wigner representation, which is
not the case of the dichotomized quadrature measurement described
above.  Considering in fact that $P(x_1^\theta,x_2^\varphi)$ can be
determined as a marginal distribution from the Wigner function one can
write from Eqs. (\ref{PH}) and (\ref{EH}):
\begin{equation}
\fl E_H(\theta,\varphi)=\int dx_1^\theta dx_2^\varphi dx_1^{\theta
+\frac\pi2} dx_2^{\varphi+\frac\pi2} sgn(x_1^\theta,x_2^\varphi) 
W(x_1^\theta,x_1^{\theta+\frac\pi2},x_2^\varphi,x_2^{\varphi+\frac\pi2})
\label{EHWPos}\;,
\end{equation} 
where the integration is performed over the whole phase-space and
without loss of generality we have considered $D_\pm={\mathbb R}^\pm$.
Eq.(\ref{EHWPos}) itself is indeed a local hidden variable description
of the correlation function, hence obeying inequality (\ref{BI2}).
\par 
In order to overcome this obstacle different strategies have been
considered by many authors, based essentially on parity measurements.
Banaszek and Wodkiewicz \cite{BW} demonstrated the nonlocality of the
TWB considering as local observable on subsystem $j$ the parity
operator on the state displaced by $\alpha_j$ (hence we will refer to
this procedure as a ``Displaced Parity (DP) nonlocality test''), which is
dichotomic and bounded:
\begin{equation}
\label{DispParity}
\Pi(\boldsymbol\alpha ) =  \otimes_{j=1}^n D_j(\alpha_j)
(-1)^{n_j} D_j^{\dagger}(\alpha_j) .
\end{equation}
In the above formula, $\boldsymbol\alpha=(\alpha_1,...,\alpha_n)$, 
while
$n_j=a^{\dagger}a$ and $D_j(\alpha_j)=exp[\alpha_j a^{\dagger}_j
-\alpha^*_ja_j]$ denote the number operator and the phase
space displacement operator for the subsystem $j$. Hence the
correlation function reads:
\begin{equation}
\label{EDP}
E_{DP}(\boldsymbol\alpha) = \langle \Pi(\boldsymbol\alpha) \rangle , 
\end{equation}
from which Bell's combinations ${\cal B}_{2,DP}$ in Eq.(\ref{BI2})
and ${\cal B}_{3,DP}$ in Eq.(\ref{BI3}) can be easily reconstructed
in the cases $n=2,3$. The reason why this procedure would be able to
reveal nonlocality also in case of quantum states characterized by a
positive Wigner function is clear using the following relation:
\begin{equation}
\label{WandP}
W(\boldsymbol\alpha) = \left(\frac{2}{\pi}\right)^n 
 \langle \Pi(\boldsymbol\alpha)  \rangle \:.
\end{equation}
Indeed, the analog of Eq.(\ref{EHWPos}) is:
\begin{equation}
E_{DP}(\boldsymbol\alpha) = \int d^{2n}\boldsymbol\lambda  
\left(\frac{2}{\pi}\right)^n W(\boldsymbol\alpha) 
\delta^{(2n)} (\boldsymbol\alpha-\boldsymbol\lambda) 
\label{EDPWPos}\; .
\end{equation} 
Being the Dirac-$\delta$ distribution unbounded, Ineqs. (\ref{BI2})
and (\ref{BI3}) are no more necessarily valid for ${\cal B}_{2,DP}$
and ${\cal B}_{3,DP}$.  The maximal violation found with this
procedure for a EPR state is ${\cal B}_{2,DP} \simeq 2.32 $
\cite{Jeong}, still far from the maximum violation admitted by quantum
mechanics.
\par
Another strategy, developed by Chen et al.~\cite{PS}, shares a similar
behavior as the one described above, allowing to reveal nonlocality for
quantum states with positive Wigner function. Interestingly, this type
of nonlocality test, which we will refer to as ``Pseudospin (PS) nonlocality
test'', admits a maximum violation for the EPR state. It can be seen as
a generalization to continuous variable systems of the one introduced
by Gisin and Peres for the case of discrete variable systems
\cite{GP}, hence, for the case of a pure bipartite system, it is equivalent
to an entanglement test \cite{Jeong}.  Let us consider the following set 
of operators, known as {\em pseudospins} in view of their commutation relations, 
${\bf s^j}=(s^j_x,s^j_y,s^j_z)$ acting on the $j$-th subsystem
\begin{eqnarray}
s^j_z &=& \sum_{n=0}^\infty\big(|2n+1\rangle_j\langle 2n+1|-|2n\rangle_j
\langle 2n|\big), \nonumber \\
s^j_x\pm s^j_y &=& 2s^j_\pm, \nonumber\\
{\bf d}^j {\bf s}^j &=& s^j_z \cos\theta^j+
\sin\theta^j(e^{i \varphi^j}s^j_-+e^{-i \varphi^j}s^j_+),
\label{PS}
\end{eqnarray}
where $s^j_-=\sum_{n =0}^\infty|2n\rangle_j\langle
2n+1|=(s^j_+)^\dagger$ and ${\bf d}^j$ is a unit vector associated to
the angles $\theta^j$ and $\varphi^j$. 
In analogy to the
spin-$\frac12$ system and defining ${\bf d}=({\bf d}^1,...,{\bf d}^n)$
the correlation function is simply given by:
\begin{equation}
 E_{PS}({\bf d})=\langle \otimes_{j=1}^n {\bf d}^j {\bf s}^j \rangle    
\label{EPS}\; ,
\end{equation}  
from which the Bell combinations ${\cal B}_{2,PS}$ and   ${\cal
B}_{3,PS}$ are evaluated. Also different representations of the
spin-$\frac12$ algebra have been discussed in the recent literature
\cite{MistaEnglert,Revzen}. In particular in Ref.~\cite{Revzen} it has 
been pointed out that different representations lead to different
expectation values of the Bell operators. Hence, the violation of Bell
inequality for continuous variable systems turns out to depend,
besides to orientational parameters, also to configurational ones.  
In the following sections we will also consider the pseudospin operators
${\bf \Pi}^j=(\Pi_x^j,\Pi_y^j,\Pi_z^j)$ taken into account in 
Ref.~\cite{Revzen}, which have the following Wigner representation:
\begin{equation}
  \label{WPi}
\fl W_{\Pi_x^j}={\rm sgn}~x_j \qquad W_{\Pi_y^j}=-\delta(x_j)~{\cal P}\frac1y_j
\qquad W_{\Pi_z^j}=-\pi\delta(x_j)\delta(y_j)  \quad,
\end{equation}
where $x_j=x_j^{0}$, $y_j=x_j^{\frac\pi2}$ and ${\cal P}$ stands for
the ``principal value''. The correlation function obtained using
operators ${\bf \Pi}^j$ will be indicated as $E_{PS}'({\bf d})=
\langle\otimes_{j=1}^n{\bf d}^j{\bf \Pi}^j\rangle$.
%%%%%%%%%%%%%%%%%%%%%%%%
\section{Three-mode nonlocality}\label{s:3m}
\par
In this section we will analyze the nonlocal properties of tripartite 
Gaussian states. In particular we will consider two classes
of states, the first one proposed by Van Loock and Braunstein
\cite{VLB}, the second one proposed in \cite{JOSAB}. The reason why we
consider this two classes is that the first is a very natural and
scalable way to produce multimode entanglement using only passive
optical elements and single squeezers, while the second one is the
simplest way to produce three mode entanglement using a single 
nonlinear optical device. Indeed, both states can be achieved
experimentally \cite{Furusawa,OL}. As concern the first class of
states, it is generated with the aid of three single mode squeezed
states combined in a ``tritter'' (a three mode generalization of a
beam-splitter). The evolution is then ruled by a sequence of single
and two mode quadratic Hamiltonians. As a consequence, being generated
from vacuum, the three-mode entangled state is Gaussian and its Wigner
function is given by:
\begin{equation}
W_S({\bf x},{\bf y})= \frac{1}{\pi^3} \exp\left[ 
-({\bf x},{\bf y}) {\bf C}^{-1} \left(
\begin{array}{c}
{\bf x} \\ 
{\bf y}
\end{array}
\right)\right]
\label{WVLB}\;,
\end{equation}
where ${\bf x}=(x_1,x_2,x_3)$, ${\bf y}=(y_1,y_2,y_3)$ are the
positions and momenta of the three modes and ${\bf C}^{-1}$ is the
inverse of the covariance matrix, whose explicit expression reads:
\begin{equation}
{\mathbf C} = \left(
\begin{array}{cccccc}
 {\cal R}  &  {\cal S} &  {\cal S}  &  0        &  0        &  0  \\
 {\cal S}  &  {\cal R} &  {\cal S}  &  0        &  0        &  0  \\
 {\cal S}  &  {\cal S} &  {\cal R}  &  0        &  0        &  0         \\
 0         &  0        &  0         &  {\cal T} & -{\cal S} & -{\cal S}  \\
 0         &  0        &  0         & -{\cal S} &  {\cal T} & -{\cal S}  \\
 0         &  0        &  0         & -{\cal S} & -{\cal S} &  {\cal T}  \\
\end{array}
\right)
\label{matC}\;,
\end{equation}
where ${\cal R}=\cosh 2r+\frac13\sinh 2r$, ${\cal T}=\cosh
2r-\frac13\sinh 2r$, ${\cal S}=-\frac43\cosh r\sinh r$ and $r$ is the
squeezing parameter (with equal squeezing in all initial modes).  The
second class of tripartite entangled states is generated in a single
non linear crystal through the following interaction Hamiltonian:
\begin{equation}
H_{int} =\gamma_1 a_1^\dag a^{\dag}_3 + \gamma_2 a_2^{\dag} a_3 + H.c.
\label{intH}\;.
\end{equation}
$H_{int}$ describes two interlinked bilinear interactions taking place
among three modes of the radiation field coupled with the support of
two parametric pumps. It can be realized in $\chi^{(2)}$ media by a
suitable configuration exposed in Ref.~\cite{OL}. Notice that the same
dynamics can be implemented in different physical systems,
including optomechanical couplers and Bose-Einstein condensates
in the linear regime \cite{CARL,Camerino,Rodionov}. The effective 
coupling constants $\gamma_j$, $j=1,2$, of the two parametric processes 
are proportional to the nonlinear susceptibilities and the pump intensities.  
If we take the vacuum $|{\bf 0}\rangle\equiv |0\rangle_1 \otimes |0\rangle_2
\otimes |0\rangle_3$ as the initial state, the evolved state 
$|{\bf T}\rangle=e^{-iH_{int}t}|{\bf 0}\rangle $ belongs to
the class of the coherent states of $SU(2,1)$ and it reads \cite{CARL,Puri}:
\begin{equation}\fl
|{\bf T}\rangle = \frac{1}{\sqrt{1+N_1}} \sum_{pq}
\left(\frac{N_2}{1+N_1}\right)^{p/2} 
\left(\frac{N_3}{1+N_1}\right)^{q/2} e^{-i(p\phi_2+q\phi_3)} 
\sqrt{\frac{(p+q)!}{p! q!}}\: |p+q,p,q\rangle
\label{T}\;,
\end{equation}
where $N_j(t)=\langle a^\dag_j(t) a(t)\rangle$ represent the average
number of photons in the $j$-th mode and $\phi_j$ are phase factors.
The explicit expressions of $N_j(t)$ are:
\begin{eqnarray}
N_2 &=& \frac{|\gamma_1|^2 |\gamma_2|^2}{\Omega ^4}
\left[\cos{\Omega t}-1 \right]^2 \;, \nonumber \\
N_3 &=& \frac{|\gamma_1|^2}{\Omega ^2} \sin^2(\Omega t) \;,
\label{Ndit}
\end{eqnarray}
with $\Omega = \sqrt{|\gamma_2|^2 -|\gamma_1|^2}$ and $N_1=N_2+N_3$.
Also for this second class, being the initial state Gaussian and the
Hamiltonian quadratic, the evolved states will be Gaussian.  The
Wigner function reads as follows \cite{JOSAB,CARLLoss}:
\begin{equation}
W_T({\bf x},{\bf y})= \frac{1}{\pi^3} \exp\left[ -({\bf x},{\bf y}) 
{\bf V}^{-1} \left(
\begin{array}{c}
{\bf x} \\ 
{\bf y}
\end{array}
\right)\right]
\label{WT}\;,
\end{equation}
where ${\bf V}^{-1}$ is the inverse of the covariance matrix, whose
explicit expression is:
\begin{equation}
{\mathbf V} = \left(
\begin{array}{cccccc}
 {\cal F}  &  {\cal A} &  {\cal B}  &  0        & -{\cal D} & -{\cal E}  \\
 {\cal A}  &  {\cal G} &  {\cal C}  & -{\cal D} &  0        &  {\cal L}  \\
 {\cal B}  &  {\cal C} &  {\cal H}  & -{\cal E} & -{\cal L} &  0         \\
 0         & -{\cal D} & -{\cal E}  &  {\cal F} &  {\cal -A} &  {\cal -B}  \\
-{\cal D}  &   0       & -{\cal L}  &  {\cal -A} &  {\cal G} &  {\cal C}  \\
-{\cal E}  &  {\cal L} &  0         &  {\cal -B} &  {\cal C} &  {\cal H}  \\
\end{array}
\right)
\label{matV}\;,
\end{equation}
where 
\begin{eqnarray}
\fl {\cal A} = 2 \sqrt{N_2(1+N_1)} \cos\phi_2 \quad & 
{\cal D} = 2 \sqrt{N_2(1+N_1)} \sin\phi_2 \quad & 
{\cal F} = 2 N_1 + 1
\nonumber \\  
\fl {\cal B} = 2 \sqrt{N_3(1+N_1)} \cos\phi_3 \quad & 
{\cal E} = 2 \sqrt{N_3(1+N_1)} \sin\phi_3 \quad &     
{\cal G} = 2 N_2 + 1
\nonumber \\                                                    
\fl {\cal C} = 2 \sqrt{N_2 N_3}\cos(\phi_2-\phi_3) \quad &
{\cal L} = 2 \sqrt{N_2 N_3}\sin(\phi_2-\phi_3) \quad & 
{\cal H} = 2 N_3 + 1
\label{Wmatrix}\;.
\end{eqnarray}
Both classes of states are fully inseparable for any value
of the coupling constants, namely cannot be written as a 
factorized state for any grouping of the modes. Therefore 
they are good candidates to reveal true tripartite 
nonlocality. Being Gaussian states, however, nonlocality 
cannot be revealed by homodyne detection. In the following
we analyze the results for displaced parity and pseudospin 
tests.
%%%%%%%%%%%%%%%%
\subsection{Displaced parity test} \label{ss:DP3}
\par
\begin{figure}[h]
  \begin{center}
    \includegraphics[width=0.5\textwidth]{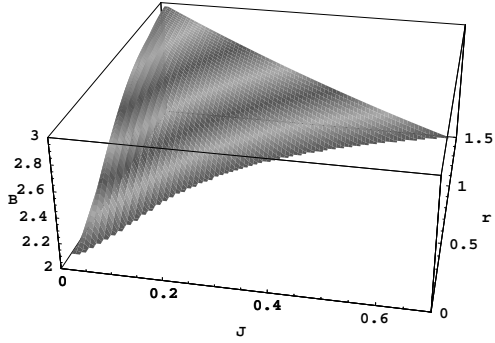}
\caption{Plot of the Bell combination (\ref{B3DPVLBGen}). 
  Only values violating Bell Inequality (\ref{BI3}) are shown.}
\label{f:B3DPVLBGen}
\end{center}
\end{figure}
\begin{figure}[t]
  \begin{center}
    \includegraphics[width=0.5\textwidth]{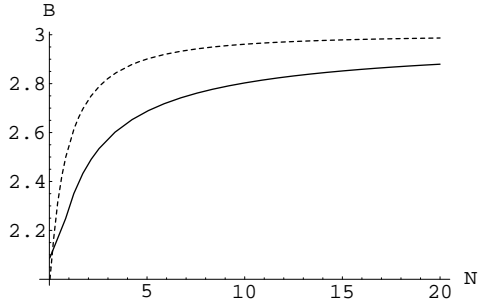}
\caption{Bell combination ${\cal B}_{3,DP}$ for state (\ref{WVLB}), 
  dotted line, and (\ref{T}), solid line. The displacement parameters
  ${\cal J}$ have been optimized to give the maximum value of ${\cal
    B}$ for $N$ fixed.}
\label{f:B3DPN}
\end{center}
\end{figure}

Let us start the study of tripartite systems nonlocality using the
``displaced parity test''.  Considering the correlation function
$E_{DP}(\boldsymbol \alpha )$ given by Eq. (\ref{EDP}), the state
(\ref{WVLB}) was found in \cite{VLB} to give a maximal violation of
${\cal B}_{3,DP} \simeq 2.32$ in the limit of large squeezing and
small displacement. The study in \cite{VLB} however was performed
for a particular choice of displacement parameters:
$\alpha_1=\alpha_2=\alpha_3=0$ and $\alpha_1'=\alpha_2'=\alpha_3'=i
\sqrt{{\cal J}}$. A numerical optimization of the displacement
parameters lead us to identify a number of parameterizations
that allow a significantly 
higher violation of Bell's inequality. As an example, consider the 
one given by $\alpha_1=\alpha_2=\alpha_3=i\sqrt{{\cal J}}$ and
$\alpha_1'=\alpha_2'=\alpha_3'=-2i \sqrt{{\cal J}}$ from which follows
that
\begin{equation}
{\cal B}_{3,DP}=3\exp \left(-12e^{-2 r} {\cal J}\right) - 
\exp\left(24 e^{2 r} {\cal J}\right)
\label{B3DPVLBGen}\;,
\end{equation}
hence the remarkably high asymptotic value of ${\cal B}_{3,DP}=3$ is
found for large $r$ and ${\cal J} \ne 0$ (see Fig.
\ref{f:B3DPVLBGen}). The importance of a suitable choice of the
displacement parameters is apparent if this asymptotic value is
compared to the violations obtained in the nonlocality study performed
in Ref.  \cite{VLB}. In that work in fact generalizations to more than
three modes of state (\ref{WVLB}) were also considered, giving an
increasing violation of Bell inequality as the number of modes
increases, but never founding a violation greater than $2.8$ .
Determining the optimal choice of the displacement parameters for a
given state is in general a challenging task. To our knowledge indeed there
exists no general prescription to find it out, and ultimately one must
rely onto a numerical analysis (see, {\em e.g.} \cite{BanaOnOff}).
Nevertheless, a careful inspection of the symmetries of the state
under consideration may be helpful. In order to clarify this observation
let us consider the explicit expression of the correlation function
$E_{DP}(\alpha_1,\alpha_2,\alpha_3)$ for state (\ref{WVLB}):
\begin{eqnarray}
\label{VLBOpt1}
\fl E(\alpha_1,\alpha_2,\alpha_3)= 
\exp
\bigg\{
-\frac{2}{3}e^{2\,r}\left[ ( y_1+y_2+y_3 )^2+(x_2-x_3)^2+(x_2-x_1)^2+(x_1-x_3)^2
\right]
\nonumber
\\
-\frac{2}{3}e^{-2\,r}\left[ ( x_1+x_2+x_3)^2+(y_2-y_3)^2+(y_2-y_1)^2+(y_1-y_3)^2
\right]  
\bigg\}
\end{eqnarray}
Hence, from Eq.~(\ref{BI3}) it follows that the Bell combination
${\cal B}_{3,DP}$ is given by the sum of three positive and one
negative term. It is reasonable to expect that the maximal violation
of nonlocality will be achieved for large $r$. We see from
Eq.~(\ref{VLBOpt1}) that, in this limit, all the correlation functions
in ${\cal B}_{3,DP}$ vanish for nonzero displacements,
unless the coefficients of $e^{2\,r}$ are zero. Hence we impose the
following system of equations, which allows for the three positive
terms in ${\cal B}_{3,DP}$ not to vanish (we consider
$\alpha_k=x_k+i\,y_k$ and $\alpha'_k=x'_k+i\,y'_k$ for $k=1,2,3$):
\begin{eqnarray}
\label{VLBOpt2}
( y'_1+y_2+y_3 )^2+(x_2-x_3)^2+(x_2-x'_1)^2+(x'_1-x_3)^2 = 0 \nonumber \\
( y_1+y'_2+y_3 )^2+(x'_2-x_3)^2+(x'_2-x_1)^2+(x_1-x_3)^2 = 0 \nonumber \\
( y_1+y_2+y'_3 )^2+(x_2-x'_3)^2+(x_2-x_1)^2+(x_1-x'_3)^2 = 0 \;.
\end{eqnarray}
We see that the parameterization used to obtain Eq.~(\ref{B3DPVLBGen}) is a solution of this system. Clearly, any other solution will give, in the limit of large $r$, the same violation given by Eq.~(\ref{B3DPVLBGen}), namely ${\cal B}_{3,DP}\rightarrow 3$. In order
to compare the violation of Bell's inequality admitted by the state
(\ref{WVLB}) with the one that will be obtained below considering the
state (\ref{T}) , it is useful to rewrite Eq.(\ref{B3DPVLBGen}) as a
function of total mean photon number $N=N_1+N_2+N_3$.  Given that
$N=3\sinh^2 r$ and optimizing the displacement ${\cal J}$ we obtain
the result shown in Fig \ref{f:B3DPN}. The asymptotic expression of
the optimized displacement as a function of $N$ is ${\cal J} =
\frac{1}{8N}{\rm ArcSinh}\left(\sqrt{\frac{N}{3}}\right)$, hence very
small angles are required.
\par
\begin{figure}[t]
  \begin{center}
    \includegraphics[width=0.5\textwidth]{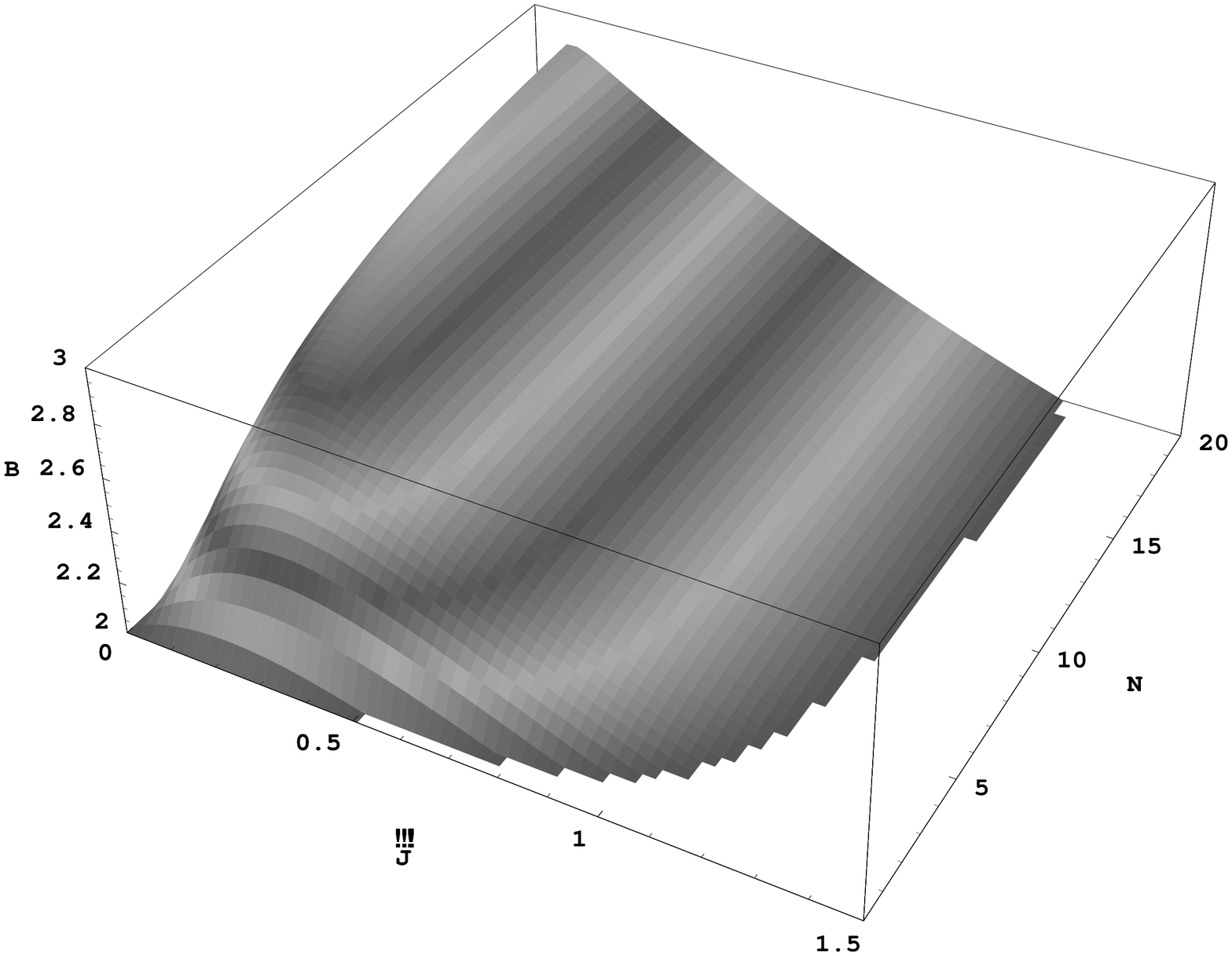}
\caption{Bell combination obtained choosing optimized displacement 
  parameters for state (\ref{T}) (see text for details). Only values
  violating Bell Inequality (\ref{BI3}) are shown.}
\label{f:B3DPT}
\end{center}
\end{figure}

Consider now the tripartite state generated by the Hamiltonian
(\ref{intH}). The correlation function is now given by Eq. (\ref{EDP})
through the Wigner function (\ref{WT}). The symmetry of the state
suggests a maximum violation of Bell inequality for
$N_2=N_3=\frac{N}{4} $ (recalling Eq.~(\ref{Ndit})), while the fact
that the separability of the state doesn't depend on the phases
$\phi_2$ and $\phi_3$ \cite{JOSAB} suggests that they are not
crucial for the nonlocality test. If we consider the same
parametrization that led us to Eq.  (\ref{B3DPVLBGen}) and fix
$\phi_2=\phi_3=\pi$, we find:
\begin{equation}
\fl {\cal B}_{3,DP}=\frac{-1 + e^{6\,{\cal J}\,\left( 1 + N + 
         2\,{\sqrt{2}}\,{\sqrt{N\,\left( 2 + N \right) }} \right)
         } + 2\,e^
      {\frac{3}{2}\,{\cal J}\,\left( 4 + 7\,N + 
            6\,{\sqrt{2}}\,{\sqrt{N\,\left( 2 + N \right) }}
            \right) }}{e^
    {4\,{\cal J}\,\left( 3 + 3\,N + 
        2\,{\sqrt{2}}\,{\sqrt{N\,\left( 2 + N \right) }} \right) 
      }}\;,
\label{B3DPTVLB}
\end{equation}
from which follows an asymptotic violation of Bell's inequalities of
${\cal B}_{3,DP} \simeq 2.89$, for large $N$ and small ${\cal J}$. A
slightly better result is found if a parametrization, more suitable
and numerically optimized for state (\ref{T}), is considered:
$\alpha_1=\frac23\sqrt{{\cal
    J}},\alpha_2=\alpha_3=\alpha_1'=0,\alpha_2'=- \sqrt{{\cal
    J}},\alpha_3'=\sqrt{{\cal J}},\phi_2=0$ and $\phi_2=\pi$. The Bell
combination ${\cal B}_{3,DP}$ for this choice of parameters is
depicted in Fig. \ref{f:B3DPT}. We note that in this case a larger
choice of angles allows the violation of Bell inequality if compared
with Fig. \ref{f:B3DPVLBGen}.
As before, optimizing the displacement ${\cal J}$ for each $N$, it is
possible to find the maximum violation of Bell inequality as a
function of $N$. We find that the asymptotic relation between the
optimized displacement and the total photon number is now ${\cal
J}N\simeq3.21$, confirming that not too small displacements are
required. The asymptotic violation of Bell's inequality is now   ${\cal
B}_{3,DP}\simeq2.99$.  To compare the results obtained for the two
states (\ref{WVLB}) and (\ref{T}) we have plotted in Fig.
\ref{f:B3DPN} the Bell combination ${\cal B}_{3,DP}$ as a function of the
mean total energy $N$, while the displacement ${\cal J}$ has been
chosen in order to maximize ${\cal B}_{3,DP}$ at fixed energy. Notice that
even if the two states have quite the same asymptotic violation, state
(\ref{WVLB}) reaches it for lower energies.
%%%%%%%%%%%%%%%%%%%%%%%%
\subsection{Pseudospin test} \label{ss:PS3}
\par
Consider now the pseudospin nonlocality test.
Let us calculate the expectation value of the correlation function
(\ref{EPS}) for the state $|T\rangle$ (for simplicity we consider
$\phi_2=\phi_3=0$). The only non vanishing contributes are given by:
        \begin{eqnarray}
  \label{c1c2c3}
\fl c_1 = \langle s_z^1 \otimes s_x^2 \otimes s_x^3 \rangle=
      \langle s_z^1 \otimes s_y^2 \otimes s_y^3 \rangle \nonumber  \\ 
       \hspace{-25pt} 
\eql 
      -\frac{\sqrt{N_2N_3}}{2(1+N_1)^2}\sum_{s,t}\left(\frac{N_2}{1+N_1}\right)^{2s}
      \left(\frac{N_3}{1+N_1}\right)^{2t}\frac{(2s+2t+1)!}{(2s)!(2t)!\sqrt{(2s+1)(2t+1)}} 
      \;, \nonumber \\ 
\fl c_2 = \langle s_x^1 \otimes s_z^2 \otimes s_x^3 \rangle =
     -\langle s_y^1 \otimes s_z^2 \otimes s_y^3 \rangle\nonumber \\ 
       \hspace{-25pt} 
     \eql
      \frac{\sqrt{N_3}}{2(1+N_1)^{3/2}}\sum_{s,t}\left(\frac{N_2}{1+N_1}\right)^{2s}
      \left(\frac{N_3}{1+N_1}\right)^{2t}\frac{(2s+2t)!}{(2s)!(2t)!}
      \sqrt{\frac{2s+2t+1}{2t+1}}
      \;, \nonumber \\
\fl c_3 = \langle s_x^1 \otimes s_x^2 \otimes s_z^3 \rangle =
     -\langle s_y^1 \otimes s_x^2 \otimes s_z^3 \rangle\nonumber \\ 
       \hspace{-25pt} 
     \eql
      \frac{\sqrt{N_2}}{2(1+N_1)^{3/2}}\sum_{s,t}\left(\frac{N_2}{1+N_1}\right)^{2s}
      \left(\frac{N_3}{1+N_1}\right)^{2t}\frac{(2s+2t)!}{(2s)!(2t)!}
      \sqrt{\frac{2s+2t+1}{2s+1}}
      \;,
\end{eqnarray}
and by $\langle s_z^1 \otimes s_z^2 \otimes s_z^3 \rangle=1 $. The
correlation function then, according to
Eqs.~(\ref{PS}) and (\ref{EPS}), reads as follows:
\begin{eqnarray}
\label{EPS3ms}
\fl E_{PS}({\bf d})=  \cos\theta^1\cos\theta^2\cos\theta^3
+c_1\cos\theta^1\sin\theta^2\sin\theta^3(\cos\varphi^2\cos\varphi^3+
\sin\varphi^2\sin\varphi^3) \nonumber \\
\lo+c_2\cos\theta^2\sin\theta^1\sin\theta^3(\cos\varphi^1\cos\varphi^3-
\sin\varphi^1\sin\varphi^3) \nonumber \\
\lo+c_3\cos\theta^3\sin\theta^1\sin\theta^2(\cos\varphi^1\cos\varphi^2+
\sin\varphi^1\sin\varphi^2)                  
\end{eqnarray}
Hence, without loss of generality, we can fix for example $\varphi^1=0$ and
$\varphi^2=\varphi^3=\pi$ and look for the maximum violation of Bell
inequality (\ref{BI3}) constructed from Eq.~(\ref{EPS3ms}). Notice
that if the coefficients $c_i$ ($i=1,2,3$) were equal to one
then the maximum violation
admitted, ${\cal B}_{3,PS}=4$, should be reached. Considering
Eqs.~(\ref{c1c2c3}) two limiting cases can be studied. First, for
large $N_2$ and small $N_3$ (or vice-versa) a numerical evaluation of the coefficients
$c_i$ shows that $c_3 \rightarrow 1$, while the other two vanish.
Hence, considering $\theta^3=0$, the correlation function (\ref{EPS3ms}) 
reduces to that of a TWB subjected to a pseudospin
nonlocality test (see Eq.~(\ref{ETWBPS}) below), hence allowing an
asymptotic violation of ${\cal B}_{3,PS}=2\sqrt2$. 
This result should be expected, since in this limiting case the state 
(\ref{T}) reduces to a TWB for modes $a_1$ and $a_2$, 
while mode $a_3$ remains in the vacuum state and factors out.
Consider now the case in
which $N_2=N_3=\frac N4$. A numerical evaluation shows that the coefficients
$c_i \rightarrow \frac12$ for large $N$, hence also in this case the maximum
violation cannot be attained. The asymptotic violation turns out to be
${\cal B}_{3,PS} \simeq 2.63$.
\par
\begin{figure}[h!]
  \begin{center}
  \includegraphics[width=0.5\textwidth]{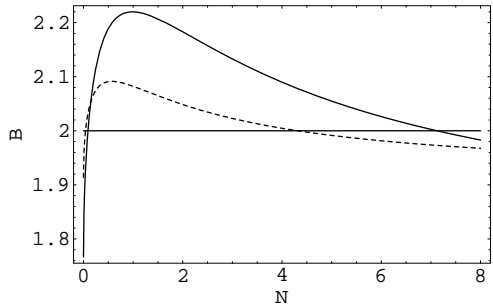}  
    \caption{Plot of Bell combination ${\cal B}_{3,PS}$ for states 
    (\ref{WT}) and (\ref{WVLB}), solid and dashed line respectively. 
    $N$ is the total number of photons as in Fig.~\ref{f:B3DPN} }
    \label{f:B3PS}
  \end{center}
\end{figure}

As already mentioned in Sec.~\ref{s:rev} other representations for
the pseudospin operators can be considered. Using Eqs.~(\ref{WPi}) and
(\ref{WT}) it is possible to calculate the correlation function
$E_{PS}'({\bf d})$. Setting again the azimuthal angles $\varphi^i=0$, the
latter shows the same structure as $E_{PS}({\bf d})$ where now the coefficient
$c_i$ are replaced by:
\begin{eqnarray}
  \label{c1c2c3primes}
  c_1'=\frac{2\arctan\left(\frac{N}{2\sqrt{1+N}}\right)}{\pi(1+N)} \qquad 
  c_2'=c_3'=\frac{2\arctan \sqrt{N}}{\pi(1+\frac N2)} \;.
\end{eqnarray}
An appropriate choice of angles leads to a maximal violation of Bell's
inequality given by ${\cal B}_{3,PS} \simeq 2.22$ (see Fig.~\ref{f:B3PS}), which is now reached
for $N \simeq 1$, value for which the coefficients $c_i'$ are
approximately near their maxima. As already pointed out, we see that
different representations of the pseudospin operators give rise to different
expectation values for the Bell operator.
\par
Applying now the same procedure to state (\ref{WVLB}) we find the same
structure for the correlation function $E_{PS}'$, where the
coefficients are now given by:
\begin{eqnarray}
  \label{c1c2c3primesVL}
  c_1'=c_2'=c_3'=\frac{-6\,\arctan (\frac{4\,\cosh (r)\,\sinh (r)}
      {\,{\sqrt{3(2 + e^{4\,r})}}})}{\pi \,
    {\sqrt{5 + 4\,\cosh (4\,r)}}}
   \;.
\end{eqnarray}
After an optimization of the angles $\theta^i$ we obtain a maximal
violation of ${\cal B}_{3,PS} \simeq 2.09$ (see Fig.~\ref{f:B3PS}), 
for $r \simeq 0.42$ ($N\simeq0.56 $) that maximizes the coefficients $c_i$.
%%%%%%%%%%%%%%%%%%%%%%%%%%%%%%%%%%
\section{Degaussified state and two mode nonlocality}\label{s:2m}
\par
In this section we consider the tripartite state (\ref{T}) as a source
of two-mode states. In particular, we study the nonlocality of a
two-mode non-Gaussian state obtained by a conditional measurement
performed on state (\ref{T}).  Gaussian states are at the heart of
quantum information processing with continuous variables. The reason
for this is that Gaussian is the character of the vacuum state of
quantum electrodynamics. This observation, in combination with the
fact that the quantum evolutions achievable with current technology
are described by Hamiltonian operators at most bilinear in the quantum
fields, accounts for the fact that the states commonly produced in
laboratories are Gaussian.  In fact, bilinear evolutions preserve the
Gaussian character of the vacuum state. In addition, it is worth
noticing that the operation of tracing out a mode from a multipartite
Gaussian state preserves the Gaussian character too, and the same
observation is valid when the evolution of a state in a standard noisy
channel is considered. Indeed, the only feasible way to ``degaussify''
a state is through a conditional measurement, or by statistically
mixing it with another Gaussian state. The reason to study
non-Gaussian states is that when the Gaussian character is lost, then
immediately the Wigner function of the state becomes negative, for
pure states, hence stronger nonclassical properties should emerge.
Actually, various authors have recently investigated the nonlocality
properties of two-mode non-Gaussian states. In particular, a twin-beam
state subjected to inconclusive photon subtraction (IPS state) has
been considered in Refs.~\cite{NhaGPatron} and \cite{IPSNonLoc}, while
in Ref.~\cite{Jeong} it has been pointed out that if the entangled
coherent states \cite{San92} could be produced experimentally they
would allow for the maximal violation ({\em i.e.}, ${\cal
  B}_2=2\sqrt2$) both in case of a DP test as well as a PS test.
\par
The most natural way to obtain a
non-Gaussian state from a Gaussian one is by elimination of its vacuum
component. In fact, such a state is necessarily described by a
negative Wigner function (in fact $\langle0|\varrho|0\rangle\propto
\int d^{2n}{\boldsymbol \alpha}W({\boldsymbol
\alpha})e^{-2|{\boldsymbol \alpha}|^2} $). Do to the structure of
state (\ref{T}) its vacuum component can be subtracted by a
conditional measurement on mode $a_3$, the same observation being valid
for mode $a_2$. Consider a photodetector able to distinguish only the
presence or the lack of photons, {\em i.e.}, an ON/OFF photodetector,
and the state $\varrho_1$ conditioned to the presence of at least one
photon. The probability operator measure (POVM) is two-valued
$\{\Pi_0,\Pi_1\}$, $\Pi_0+\Pi_1={\bf I}$, with the element associated
to the ``no photons'' result given by
\begin{equation}
\label{NoPh} \Pi_0= \hbox{\bf I}_1 \otimes \hbox{\bf I}_2
\otimes \sum_{n} (1-\eta)^n |n\rangle_3{}_3\langle n| \:,
\end{equation}
where $\eta$ is the efficiency of the photodetector.
The probability of the outcome is given by
\begin{eqnarray}
  \label{P1} P_1 = \hbox{Tr}_{123} \left[|{\bf T}\rangle\langle{\bf T}|\:
  \Pi_1\right] = \frac{\eta N_3}{(1+\eta N_3)} \:,
\end{eqnarray}
while the conditional output state reads as follows
\begin{eqnarray}
  \fl \label{rho1} \varrho_1 = \frac{1}{P_1}
  \hbox{Tr}_3\left[|{\bf T}\rangle\langle{\bf T}|\:
  \Pi_1\right] \nonumber \\ 
  \hspace{-50pt}
  \eql \frac{1+\eta N_3}{(1+N_1)\eta N_3} \sum_{p=1}^\infty
  \left(\frac{N_3}{1+N_1}\right)^p \frac{1-(1-\eta)^p}{p!}(a^\dagger)^p
  \sum_{n,n'} \left(\frac{N_2}{1+N_1}\right)^{n+n'} |n\,n\rangle
  \langle n'\,n'| a^p \nonumber \\ 
  \hspace{-50pt}\eql
  \frac{1+\eta N_3}{(1+N_1+N_2)\eta N_3} \sum_{p=1}^\infty
  \left(\frac{N_3}{1+N_1}\right)^p \frac{1-(1-\eta)^p}{p!} 
  (a^\dagger)^p |X\rangle\langle X| a^p 
  \:,
\end{eqnarray}
where we have identified the TWB with
$X=\sqrt{\frac{N_2}{1+N_1}}$.  To calculate the Wigner function
associated with state $\rho_1$, consider that the characteristic
function of the POVM $\Pi_1$ reads as follows:
\begin{equation}
\label{Pi1Chi}
 \chi[\Pi_1](\mu)=\pi\delta^2(\mu)-\frac1\eta\exp\left[-|\mu|^2
\frac{2-\eta}{2\eta}\right]\;,  
\end{equation}
hence the characteristic function of $\varrho_1$ is given by:
\begin{equation}
\label{Rho1Chi}
\fl\chi[\varrho_1](\lambda_1,\lambda_2)=\frac{1}{P_1}
\left\{
\chi[|T\rangle\langle T|](\lambda_1,\lambda_2,0)
-\frac1\eta\int \frac{d^2\mu}{\pi}\chi[|T\rangle\langle T|](\lambda_1,\lambda_2,\mu)
\exp\left[-|\mu|^2\frac{2-\eta}{2\eta}\right]
\right\}
\;.  
\end{equation}
After some algebra the Wigner function associated with state $\rho_1$
can now be calculated. It reads as follows:
\begin{eqnarray}
\fl W_1({\bf x},{\bf y})= 
\frac{1+\eta N_3}{4\eta N_3}
\Bigg\lbrace\left(\frac{2}{\pi}\right)^2\frac{1}{\sqrt{\det V'}}
\exp\left[ -({\bf x},{\bf y}) \left({\bf V'}\right)^{-1} \left(
\begin{array}{c}
{\bf x} \\ 
{\bf y}
\end{array}
\right)\right] \nonumber \\
\lo- \frac{1}{\eta}
\left(\frac{2}{\pi}\right)^2\frac{2}{\sqrt{\det D}}
\exp\left[ -({\bf x},{\bf y}) \left({\bf D}^{-1}\right)' \left(
\begin{array}{c}
{\bf x} \\ 
{\bf y}
\end{array}
\right)\right]
\Bigg\rbrace
\label{W1}\;,
\end{eqnarray}
where, from now on, ${\bf x}=(x_1,x_2)$, ${\bf y}=(y_1,y_2)$, and ${\bf D}={\bf V}+{\rm diag}
\big(0,0,\frac{2-\eta}{\eta},0,0,\frac{2-\eta}{\eta} \big)$.
In order to simplify the notation we have indicated with ${\bf O}'$ 
the $4\times4$ matrix obtained from the $6\times6$ matrix ${\bf O}$ 
deleting the elements corresponding to the third mode (3-th row/column 
and 6-th row/column), due to the trace over the 3-th mode.
Of course, the easiest way to obtain a bipartite state from state
(\ref{T}) is to discard a mode, say the third one, by tracing over it.
The state $\varrho_{Tr}$ then obtained is simply given by the
following Wigner function:
\begin{equation}
W_{Tr}({\bf x},{\bf y})= 
\left(\frac{2}{\pi}\right)^2\frac{1}{4\sqrt{\det V'}}
\exp\left[ -({\bf x},{\bf y}) \left({\bf V'}\right)^{-1} \left(
\begin{array}{c}
{\bf x} \\ 
{\bf y}
\end{array}
\right)\right]
\label{Wtraced}\;.
\end{equation}
Being the Wigner function $W_{Tr}$ Gaussian, we expect that this state
will exhibit weaker nonlocality with respect to state (\ref{rho1}). In
the rest of the section the nonlocal properties of the usual TWB
state and of the states (\ref{rho1}) and (\ref{Wtraced}) will be
compared. Notice that state (\ref{WVLB}) can be considered as an
extension to three modes of the TWB. All the three nonlocality
tests introduced in Sec.~\ref{s:rev} will be taken into account.
%%%%%%%%%%%%%%%%%%%%%%%%
\subsection{Displaced parity test} \label{ss:DP2}
\par
\begin{figure}[h]
  \begin{center}
    \includegraphics[width=0.5\textwidth]{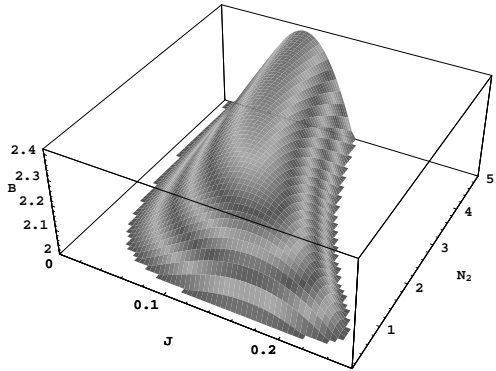}
\caption{Bell combination obtained choosing optimized displacement 
  parameters for state $\varrho_1$ (see text for details). Only values
  violating inequality (\ref{BI2}) are shown.}
\label{f:B2DPTWBA}
\end{center}
\end{figure}

We first study the violation of inequality (\ref{BI2}) in the case of
a ``displaced parity test''. As already mentioned, in Ref.~\cite{BW}
Banaszek and Wodkiewicz found for the first time that the TWB
state exhibit a violation of local realism. They obtained the
following asymptotic violation for infinite energy: ${\cal B}_{2,DP}
\simeq 2.19$. Generalizing their procedure this result can be
improved, yielding to a maximum asymptotic violation of ${\cal
  B}_{2,DP} \simeq 2.32$ \cite{Jeong}. We have considered the
following parametrization to obtain the maximum violation for a
TWB: $\alpha_1=-\alpha_2=i\sqrt{{\cal J}}$ and
$\alpha_1'=-\alpha_2'=-3i \sqrt{{\cal J}}$. The asymptotic relation
between the squeezing parameter and the displacement angles is
$e^{2r}{\cal J}=\frac{\log 3}{32}$.  Using the same parametrization
and considering the Bell combination ${\cal B}_{2,DP}$ for the state
$\varrho_{Tr}$, it turns out that the same asymptotic value of the
TWB is reached for large $N_2$ and small $N_3$.  In fact, as 
already noticed, when this
limit is considered the original tripartite state (\ref{T}) reduces to a factorized 
state composed by a TWB and the vacuum state.
Consider now the conditional state $\rho_1$ and again the case of
large $N_2$ and small $N_3$, say $N_3=10^{-2}\frac{1}{N_2}$. As in the
tripartite case the phase coefficients $\phi_2$ and $\phi_3$ play no
rule in the characterization of nonlocality. A stronger violation of
Bell inequality is then found and it is depicted in Fig.
(\ref{f:B2DPTWBA}), where the parametrization
$\alpha_1=\frac12\alpha_2=\frac13\alpha_1'=i\sqrt{{\cal J}}$ and
$\alpha_2'=0$ has been adopted.
Indeed the asymptotic violation is higher then the previous, namely
${\cal B}_{2DP} \simeq 2.41$. It can be found, for large $N_2$, when
${\cal J} N_2 \simeq 0.042$. A comparison with the violation of
nonlocality attained with a IPS state shows an identical asymptotic
behavior \cite{IPSNonLoc}. Nevertheless, the scheme proposed here
offers the advantage that the production rate of state $\varrho_1$
[{\em i.e.}, the conditional probability (\ref{P1})] is much greater
then the production rate of IPS state [see Ref.~\cite{IPSTele},
Eq.(14)]. This is due to the fact that only a single ON/OFF detection
is required to produce $\varrho_1$, rather than the coincidence of two
ON/OFF detections for the case of IPS state. This could be useful from
a practical viewpoint.
%%%%%%%%%%%%%%%%%%%%%%%%
\subsection{Pseudospin test}  \label{ss:PS2}
\par
\begin{figure}[h]
  \begin{center}
    \includegraphics[width=0.5\textwidth]{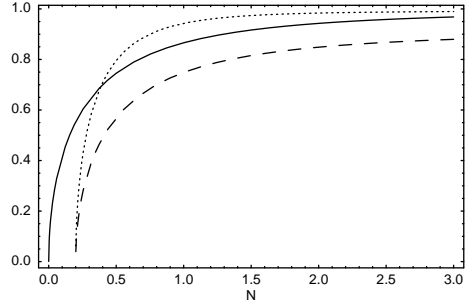}
\caption{Comparison between the values of the functions $f_{TWB}$ 
  (solid line), $f1$ (dotted line) and $f_{Tr}$ (dashed line) defined
  in the text (the summation has been numerically performed for $\eta
  =0.8$ and $N_3=0.1$).}
\label{f:B2PS}
\end{center}
\end{figure}

Let us now focus on the ``pseudospin nonlocality test''. Considering a
TWB state, it is known that the correlation function (\ref{EPS})
has the following expression (setting to zero the azimuthal angles)
\cite{PS}:
\begin{equation}
\label{ETWBPS}
E_{PS}(\theta_1,\theta_2)=\cos\theta_1\cos\theta_2+f_{TWB}
\sin\theta_1\sin\theta_2 \:,
\end{equation}
where, denoting with $N$ the total photon number,
\begin{equation}
\label{FTWB}
f_{TWB}=\frac{\sqrt{N(N+2)}}{1+N} \:.
\end{equation}
It turns out that the violation of Bell inequality in this contest
increases monotonically to the maximum value of $2\sqrt2$ as the
function $f_{TWB}$ goes to unity. A straightforward calculation shows
that an expression identical in form to Eq.~(\ref{ETWBPS}) can be
found both in case of state $\varrho_1$ and $\varrho_{Tr}$, where the
following functions $f_1$ and $f_{Tr}$ can be identified:
\begin{eqnarray}
\label{F1FTr}
\fl f_1=2\,{\sqrt{\frac{N_2}
        {1 + N_1}}}\,\frac{
    \left( 1 + N_3\,\eta  \right)}{N_3\,
    \left( 1 + N_1 \right) \,\eta } \,
    \sum_{k,p = 0}^{\infty }
         \frac{\left( 2\,k + p \right) !}
           {\left( 2\,k \right) !\,p!}\,
          {\sqrt{\frac{2\,k + p + 1}{2\,k + 1}}}\,
          \left( 1 - {\left( 1 - \eta  \right) }^p \right) \,
          {\left( \frac{N_3}
              {1 + N_1} \right) }^p\,
          {\left( \frac{N_2}
              {1 + N_1} \right) }^
           {2\,k} 
 \:, \nonumber \\
\fl f_{Tr}=2\,{\sqrt{\frac{N_2}
        {1 + N_1}}} \frac{1}{1 + N_1} \,
       \sum_{p,q = 0}^{\infty }
         {\left( \frac{N_2}
              {1 + N_1} \right) }^
           {2\,p}\,{\left( \frac{N_3}
              {1 + N_1} \right) }^
           {2\,q}\,\frac{\left( 2\,p + 2\,q \right) !}
           {\left( 2\,p \right) !\,\left( 2\,q \right) !}\,
          {\sqrt{\frac{2\,q + 2\,p + 1}{2\,p + 1}}} 
     \:.
\end{eqnarray}
In order to compare the violations in the three different cases, let
us fix as in the previous subsection a small value for $N_3$. A plot
of the functions $f_{TWB}$, $f_1$ and $f_{Tr}$ versus the total number
of photons of the TWB for the former and of the initial three-partite
state for the latter two is given in Fig.~(\ref{f:B2PS}).  It can be
seen that state $\varrho_1$ achives large violations for smaller
energies with respect to the other two states. Finally, a comparison
with the violation attained with the IPS state may be found in
Ref.~\cite{napoli}.
%%%%%%%%%%%%%%%%%%%%%%%%
\subsection{Homodyne detection} \label{ss:H2}
\par
\begin{figure}[h]
\begin{center}
  \includegraphics[width=0.5\textwidth]{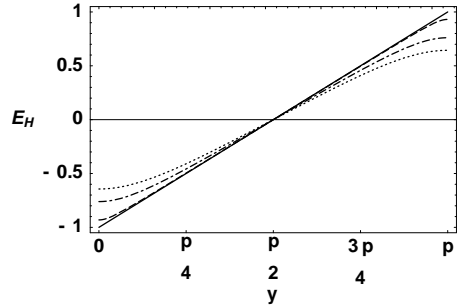}
\caption{Comparison between the correlation functions obtained from 
  two spin-$\frac12$ particles classically correlated (solid line) and
  from Eq.~(\ref{EHrho1}): $N_2=0.5$ (dotted line),$N_2=1$ (dot-dashed
  line),$N_2=5$ (dashed line). In all cases we have fixed $N_3=0.5$
  and $\eta=1$.}
\end{center}
\label{f:E2H}
\end{figure}

The negativity of the Wigner function (\ref{W1}) may suggest to
perform a nonlocality test based upon a Homodyne detection scheme.
While the positivity of a Wigner function avoid to violate Bell
inequality (\ref{BI2}) with such a test, its negativity is yet not
sufficient in general to ensure a violation. Quantum states with
negative Wigner function that doesn't violate local realism with a
homodyne test are given for example in Refs.~\cite{MunroMancini}.
Considering state (\ref{rho1}) it is necessary to calculate the
correlation function (\ref{EHWPos}). Substituting the Wigner function
(\ref{W1}) into Eq.~(\ref{EHWPos}) and performing the integral we
obtain the following result:
\begin{eqnarray}
\fl E_H(\psi)= 
\frac{1+\eta N_3}{4\eta N_3}
\Bigg\{
-\left(\frac{2}{\pi}\right)^2\frac{1}{\sqrt{\det V'}}
\left[2\,\left( 1 + 2\,N_3 \right) \,\pi \,
  \arctan \left(\frac{2\,\cos\psi }
    {{\sqrt{\frac{\left( 1 + 2\,N_1 \right) \,\left( 1 + 2\,N_2 \right) }
          {\left( 1 + N_1 \right) \,N_2} - 4\,\cos^2\psi}}}\right)\right]
\nonumber \\
\fl -\frac{1}{\eta}
\left(\frac{2}{\pi}\right)^2\frac{2}{\sqrt{\det D}}
\left[
\frac{2\,\pi \,\left( -1 + N_3\,\left( -2 + \eta \right)  \right)}{1 + 
    N_3\,\eta} \,
    \arctan \left(\frac{2\,\cos\psi}
      {{\sqrt{\frac{\left( 1 + 2\,N_1 - N_3\,\eta \right) \,
              \left( 1 + 2\,N_2 + N_3\,\eta \right) }{\left( 1 + 
                N_1 \right) \,N_2} - 4\,\cos^2\psi}}}\right)
\right]
\Bigg\}
\label{EHrho1}\;,
\end{eqnarray}
where $\psi=\theta+\varphi+\phi_2$.  A plot of the correlation
function (\ref{EHrho1}) is depicted in Fig.~(\ref{f:E2H}) for unitary
efficiency $\eta$, together with the classical correlation function of
two spin-$\frac12$ particles \cite{Peres} (see caption for details).
Unfortunately, the comparison shows clearly that the correlation given by
Eq.~(\ref{EHrho1}) is always lower then the classical one, hence
despite the negativity of Wigner function (\ref{W1}) no violation of
Bell inequality is achievable with this scheme.
%%%%%%%%%%%%%%%%%%%%%%%%
\section{Conclusions} \label{s:conclusions}
A detailed analysis of the nonlocality properties of multipartite  
continuous variables systems obtained by parametric optical 
systems has been presented, using the more recent approaches developed 
to this aim.  We have considered in particular two
classes of tripartite Gaussian states that seems promising for quantum
communication purposes in order to implement multipartite quantum
protocols. The results show that for these states a nonlocality test
based on displaced parity measurements is more suitable to reveal
violation of local realism than one based on pseudospin operators.
This results are just the opposite of what have been obtained 
for the bipartite case. Notice, however, that a systematic approach to
pseudospin operators for continuous variables haven't been developed
yet, hence we have only used the two inequivalent configurational
parameterizations more suitable for calculations. 
For displaced parity test we obtained a remarkably high asymptotic value for the 
Bell parameter, ${\cal B}_{3,DP}\simeq3$. In this case the choice of a 
proper parametrization, suitable for the state under investigation, have 
revealed crucial. 
\par 
We have also explored the possibility to enhance nonlocality
in bipartite systems considering states endowed with a nonpositive
Wigner function. We investigated a method to
conditionally produce such a state from the tripartite systems
considered above. As expected, the Bell parameter reaches a value
higher than for a TWB, namely ${\cal B}_{2,DP}\simeq2.41$. Also
in the case of a pseudospin test an enhancement of nonlocality has
been demonstrated, while a violation of local realism using a
dichotomic quadrature measurement cannot be achieved.
%%%%%%%%%%%%%%%%%%%%%%%%%%%%%%%%%%%
\section*{Acknowledgments}
The authors are grateful to S.~Olivares and M.~S.~Kim for fruitful discussions.
%%%%%%%%%%%%%%%%%%%%%%%%

%%%%%%%%%%%%%%%%%%%%%%%%%%%%%%%
\end{document}